\documentclass[aps,prb,twocolumn,superscriptaddress,showpacs]{revtex4}
\usepackage{graphicx}
\usepackage{amsmath}
\usepackage{amsfonts}
\usepackage{bm}
\usepackage{color}
\input colordvi

\newcommand{\Schroedinger}{Schr\"{o}dinger }
\newcommand{\ave}[1]{\langle #1 \rangle}

\begin{document}
\title{Dissociation of a Hubbard--Holstein bipolaron driven away from equilibrium by a constant electric field}
 
\author{D. Gole\v z}
\affiliation{J. Stefan Institute, SI-1000 Ljubljana, Slovenia}
\author{J. Bon\v{c}a}
\affiliation{J. Stefan Institute, SI-1000 Ljubljana, Slovenia}
\affiliation{Department of Physics, FMF, University of Ljubljana, Jadranska 19, SI-1000 Ljubljana, Slovenia }
\author{L. Vidmar}
\affiliation{J. Stefan Institute, SI-1000 Ljubljana, Slovenia}

\begin{abstract}
Using a variational  numerical method we compute  the time-evolution  of the Hubbard-Holstein bipolaron from its ground state  when at $t=0$ the constant electric field is switched on.  The system is evolved taking into account full quantum effects until it reaches a quasistationary state. In the zero-field limit  the current  shows Bloch oscillations characteristic for the adiabatic regime where the electric field causes  the bipolaron to evolve along the quasiparticle band. Bipolaron remains bound and the net current remains zero in this  regime. At larger electric fields the system enters the dissipative regime with a finite quasistationary current. Concomitantly, the bipolaron dissociates into two separate polarons. By examining different parameter regimes we show that the appearance of a finite quasistationary current is inevitably followed by the dissociation of the bipolaron.
\end{abstract}
	\maketitle

\section{Introduction}
Nonequilibrium phenomena in interacting many--body systems have recently drawn significant attention.
One of the most intriguing challenges is to understand how a system responds  to an external electric field and to unravel the microscopic  mechanism that gives rise to a constant current and a steady  increase of the total energy of the system.
From a theoretical point of view, a non--linear current--voltage characteristics with a threshold behavior is well--established,~\cite{oka03,kirino10,eckstein10} and recently the appearance of a regime of negative differential resistivity has been observed in several interacting systems.~\cite{tjE,holE,aron11,amaricci11}
Moreover, the onset of Bloch oscillations (well--known for the noninteracting systems) has been observed in different correlated systems at very large electric fields~\cite{eckstein10,aron11,freericks2008} and even in the integrable models.~\cite{marcin10}
Oscillations in the current response can also exhibit a more complex pattern leading to beats in a certain regime of model parameters.~\cite{freericks2008,karlsson11}
Nevertheless, the  heating arising from the  energy flow to the system represents a serious obstacle in calculation of a constant nonzero current for systems at half--filling,~\cite{marcin10,marcin11} therefore most of the calculations investigating current--voltage characteristics have been limited to one dimension.~\cite{hm10,kirino10}
One of the  widely used concepts is to couple the system to a thermal bath, which at a loss of quantum coherence enables calculations of a quasistationary current.
Recently, the influence of coupling to a thermal bath on the current--voltage characteristics of the Hubbard model has been investigated.~\cite{amaricci11} The reported influence is  similar to the role  of  quantum phonons on dissipation of the excess energy and consequently  on the current--voltage characteristics of the Holstein model.~\cite{tjholE}

A complementary approach to investigation of quantum systems out of equilibrium is to study dynamics of a single carrier propagating in dissipative medium,~\cite{trugman07,fehske11} in particular a charge carrier driven by a constant electric--field.~\cite{thornber}
Such problems have been recently solved for a driven carrier doped into Mott insulator~\cite{tjE} and a driven carrier coupled to Holstein phonons,~\cite{holE} where the nonzero current arises due to constant emission of magnons and phonons, respectively.
Recently, using a state--of--the--art truncated Lanczos method~\cite{bonca07,bonca08,vidmar09} the quasistationary current in a doped 2D Mott insulator coupled to phonons was calculated.~\cite{tjholE}
The advantage of such approaches is reflected in calculation of a stable well-defined current while preserving the full quantum nature of the problem.
In addition, the total energy gained by hopping of a charge carrier along the direction of the electric field is entirely absorbed within the microscopic model. This idea enables the formulation of the time--evolution of different subsystems without coupling to an external reservoir.

The aim of our study is to extend a recent study of a driven charge carrier coupled to Holstein phonons~\cite{holE} to a two--particle problem as described within the Hubbard--Holstein Hamiltonian. There are two sources of the particle-particle  interaction in this problem: a) an indirect interaction, mediated by the  electron--phonon coupling  and b) the direct on--site Coulomb interaction.  
This motivates us to address the following questions:
{\it (i)} Is it possible to drive a bound pair of electrons out of equilibrium in such way that a finite current is obtained without breaking the pair?
We show that a Hubbard--Holstein bipolaron at initial time $t=0$,  driven by a constant electric field,  begins to dissociate  as soon as a nonzero quasistationary current is reached.
{\it (ii)} How is the quasistationary current calculated per particle changed in a two--electron system with respect to the single--electron case?
{\it (iii)}  We analyze beats in the transient current response. We show that the beats in our model emerge if the energy scale associated with the Bloch oscillations within the quasiparticle (QP) band coincides with the energy gap between the QP band and the low--energy continuum of excited states.

The paper is organized as follows.
We introduce the model and numerical method in Sec. II. In Sec. III we show numerical results for the short--time behavior of a driven Hubbard--Holstein bipolaron with a focus on the dissociation of a bound state due to the constant external electric field.
In Sec. IV we address properties of the current--field characteristics in the quasistationary state while in Sec. V we analyze the Fourier spectrum of the real--time current. We give conclusions in Sec. VI.

\section{Model and numerical method}
We define the one-dimensional time-dependent Hubbard--Holstein Hamiltonian, threaded by an external flux
\begin{eqnarray}
\vspace*{-0.0cm}
H &=& -t_0 \sum_{{l} ,\sigma} \left[ {\mathrm e}^{i \phi(t)}\; \tilde{c}^{\dagger}_{{l}, \sigma}\tilde{c}_{{l+1}, \sigma} + {\mathrm H.c.} \right] + {g} \sum_{{j}} n_{{j}} (a_{{j}}^\dagger + a_{{j}}) \nonumber \\
 & + & \omega_0\sum_{{j}} a_{{j}}^\dagger  a_{{j}}+ U \sum_{{j}} n_{j\uparrow} n_{j\downarrow}, \label{ham}
\end{eqnarray}
where $c_{j}^{\dagger}$ and $a_{j}^{\dagger}$ are local electron and phonon creation operators on site $j$, respectively, and $n_{ j}=\sum_\sigma c_{{j},\sigma}^{\dagger}c_{{j}\sigma}$ is electron density. $\omega_{0}$ denotes the dispersionless optical phonon frequency, $t_{0}$ is the nearest--neighbor hopping amplitude and $U$ represent Hubbard repulsion between on-site electrons.
We set $\omega_0/t_0=1.0$ throughout the paper.
The dimensionless electron-phonon (EP) coupling strength is defined as $\lambda=g^{2}/2 t_{0} \omega_{0}.$ The electric field is applied through a time--dependent Aharonov-Bohm flux 
$\phi(t)= - F t$, which generates force on the charged particle due to the  Faraday's law. The constant electric field $F$ is switched on at time $t=0$ and is measured in units of $[t_{0}/e_{0}a]$, where $e_{0}$ is the unit charge and $a$ is lattice constant. The time is furthermore measured in units of $[t_{0}/\hbar]$, and we set $t_0=\hbar=e_0=a=1$.

In our calculations we obtain reliable numerical results for  electric fields in the range between  $10^{6}-10^{7} V/cm$, if the model parameters are set to  $t_{0}=0.1eV$ and $a=10^{-10}m$.
These fields become relevant for comparison with experiments as presented in Ref.\cite{oka2009a} if the extrapolation to zero temperature $T\rightarrow 0$ is taken.

We  solve the time-dependent \Schroedinger equation for two electrons coupled to phonon degrees of freedom, with the use of the numerical method based on the exact diagonalization of the variational Hilbert space (VHS) that led to numerically exact solutions of the polaron and bipolaron ground and low-lying excited-state properties.\cite{bonca99,bonca00,vidmar10}
In the case of the Hubbard--Holstein model the method generates the VHS starting with the translationally invariant initial state $|\varphi_0 \rangle$ where both electrons are located on the same site with no phonon degrees of freedom. 
The VHS is then generated by repeatedly applying the off-diagonal terms of Hamiltonian in Eq.~(\ref{ham}),
\begin{equation}
\left\{|\varphi_{l}^{N_h,M} \rangle \right\}=
[H_{kin}  + \sum_{m=1}^{M} \left(H_{g} \right)^{m}]^{N_h}
|\varphi_0 \rangle,
\label{Eqn:gener}
\end{equation}
where $H_{kin}$ and $H_{g}$ are the  first and the second term of the Hamiltonian in Eq.~(\ref{ham}). Parameters $N_{h}$ and $M$ determine the size of the VHS. The parameter $M > 1$ ensures good convergence in the strong EP coupling regime.
To reach the intermediate or weak coupling regime $\lambda < 1$, we introduced an additional parameter $N_{phmax}$ which limits the maximal number of phonon quanta, enabling larger values of $N_{h}$.

We first calculate the ground state at $F=0$ of the Hubbard--Holstein Hamiltonian in Eq.~(\ref{ham}).
Equilibrium properties of bipolarons have been extensively studied in the literature,~\cite{bonca00,bonca00a,macridin04,hohenadler05a,hohenadler05b,hague07,berciu07,eagles07,devreese09,osor10} with particular emphasis on the conditions for formation of a bound bipolaron state.
We switch on the uniform electric field $F$ at $t=0$ and start the time propagation using the time-dependent Lanczos technique.~\cite{lantime} 
The time step is taken to be small enough ($dt/t_{B}=0.5 \times 10^{-3}$ where
$t_{B}=2 \pi/F$ denotes the Bloch oscillation period) to ensure good numerical accuracy. 
We investigate only the $S_{tot}^{z}=0$ subspace. 
An important observable is the time--dependent average of the current operator  per particle, $j(t)=\langle \hat{I}(t) \rangle/2$, where 
\begin{equation}
  \hat{I}(t)=i\left( \sum_{l,\sigma} e^{-i F t} c_{l,\sigma}^{\dagger} c_{l+1,\sigma} - {\rm H.c.} \right).
\end{equation}
 Since we are dealing with the time--independent field $F$, the time integral of the current is directly related to a change of the total energy 
 \begin{equation}
   \int_{0}^{t}j(t')dt'= \Delta h(t)/2F,	
   \label{Eqn:sum_rule}
 \end{equation}
where $\Delta h(t) = \langle H(t) \rangle - \langle H(t=0) \rangle$.\cite{holE}
We also calculate the average number of phonon quanta at time $t$, $\langle n_{ph} \rangle(t) = \langle \sum_j a_{{j}}^\dagger  a_{{j}} \rangle$.

\begin{figure}
\includegraphics[width=0.45\textwidth]{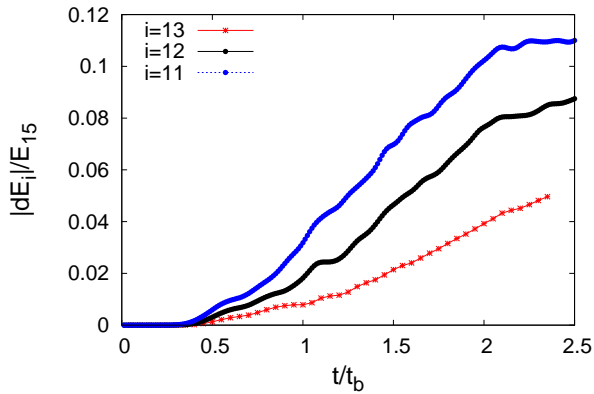}
\caption{(Color online) Time dependence of the relative difference between energies for different systems sizes $|E_{N_h}-E_{15}|/E_{15}$, where $E_{N_h}$ is energy of the system generated by the functional generator, see Eq.~(\ref{Eqn:gener}), using $N_h$ as parameter and maximal number of phonons were limited to $N_{phmax}=15$.
We set $\lambda=0.9$ and $F=1.0$ in this figure.}
\label{fig8}
\end{figure}

Due to the  finite size of the VHS we are able to run the propagation only until the system reaches  the boundary of the variational space.
There are two typical boundaries the system can reach: {\it (i)} the size of the bipolaron  becomes larger than $N_h$,  or {\it (ii)} the total number of phonons in the system approaches  the maximal allowed value $N_{phmax}$.  We checked the finite size effects by comparing the expectation values of different observables for different system's sizes.
In Fig.~\ref{fig8} we represent the time dependence of the relative difference between energies for different systems sizes, namely $|E_{i}-E_{N_h=15}|/E_{N_h=15}$, where $i$ represents the parameter $N_h$ of the functional generator in Eq.~(\ref{Eqn:gener}) and the maximum number of phonons were limited to $N_{phmax}=15$. Our simulations were stopped  when the relative energy difference between largest and second largest system was more that $5\%$.
Using the sum rule, Eq.~(\ref{Eqn:sum_rule}), we were able to independently control the precision of the time-evolution, where errors occur due to time discretization. 
\newline

\section{ Real--time behavior}
Previous studies\cite{oka03,holE,fishman} have shown the importance of the low-energy spectrum for the  short--time behavior of the interacting systems after switching on the  static electric field.  In terms of short--time dynamics there exist at least three  distinct  energy scales. First is the binding energy of the bipolaron $\delta$ representing the energy required to dissociate a bound bipolaron into two separate polarons, $\delta=E_{bi}-2E_{pol}$, where $E_{bi}$ and $E_{pol}$ are the energies of a bipolaron and a polaron, respectively. 
The second is the phonon frequency $\omega_0$, the third is the gap in the excitation spectrum $\Delta$ separating the top of the bipolaron band from either the continuum of states or the excited bipolaron band. 
Due to the existence of a finite gap, there exists a threshold field $F_{th}$ separating the adiabatic regime for $F< F_{th}$ and a dissipative regime with a finite current at  $F> F_{th}$, see also Refs.~\cite{oka03,holE}. 
Characteristic features of the adiabatic regime are periodic Bloch oscillations as the bipolaron moves along the bipolaron band. 
As a consequence, a time--averaged current in the adiabatic regime is zero.
On the other hand, in the dissipative regime a quasistationary state is reached after a short transient time $t>t_{tr}$.
A quasistationary state is characterized by a constant time--independent current  $j(t)=\bar\jmath$ and  a linear growth of the total energy of the system, 
\begin{equation} \label{hQS}
\Delta \dot h(t)=\bar\jmath \; 2F.
\end{equation} 
In the case of a driven polaron it has been shown that the steady increase of $\Delta h(t)$ is entirely due to the increase of phonon excitations left in the wake of the traveling polaron.\cite{holE}
 
A reasonable conjecture in the case of the  bipolaron is that for a large binding energy, {\it i.e.}, for 
 $\delta > \omega_{0}$, the  system on short time scale propagates as a bound bipolaron, depositing the excess energy in the form of excited  phonons left in its wake.
Our calculations show that in the weak coupling regime ($\lambda\ll 1$) the bipolaron dissociates on a time scale much shorter than $t_B$. 
We therefore focus in the following on the  intermediate and strong coupling regimes ($\lambda=0.5$ and $0.9$, respectively), where the propagation of a bound pair  under the electric field  is more likely.
However, as we shall show, such propagation  is not observed in our calculations, instead, the bipolaron begins to dissociate as soon as the system enters the dissipative regime with a finite electric current. 
 
In this section we focus on $U=0$. The motivation for choosing  $\lambda=0.5$ and $\lambda=0.9$ is that  the binding energy  is smaller than $\omega_0$ ($\delta \approx 0.58<\omega_0$) for $\lambda=0.5$ while  for $\lambda=0.9$ it is much larger than $\omega_0$ ($\delta\approx 2.04>\omega_0$).
We also note that in the case of $\lambda=0.9$ there exists an excited bound state at $\Delta\sim0.85$ above the ground state at $k=0$. 
This state represents a bound state of the  bipolaron with excited phonon cloud. Similar bound excited states  have  been well-studied for the case of the Holstein polaron.~\cite{osor06,vidmar10}
They form dispersive bands just below the one-phonon continuum that starts at $\omega_0$ above the ground state energy. 

\begin{figure}
\includegraphics[width=0.40\textwidth,height=0.45\textwidth]{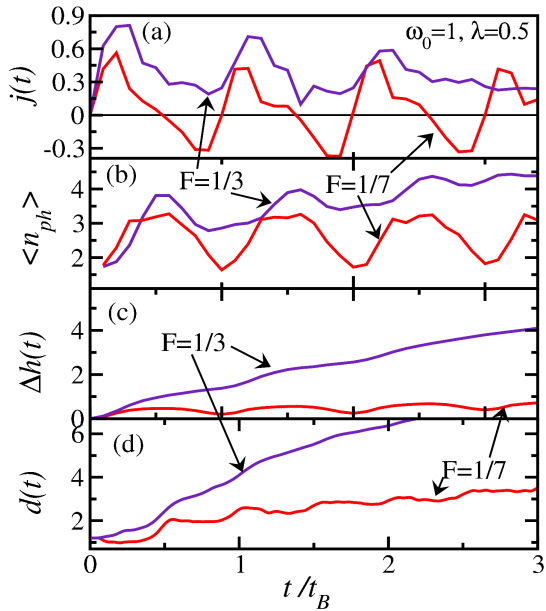}
\caption{(Color online) $j(t)$, $\langle n_{ph} \rangle(t)$, $\Delta h(t)$ and $d(t)$ vs $t/t_{B}$ for $\lambda=0.5$. We represent characteristic fields in nearly adiabatic regime and when $F \gtrsim F_{th}$, {\it i.e.}, $F=1/7$ and $F=1/3$. The accuracy of the propagation was checked by the comparison to the energy-gain sum rule, Eq.~(\ref{Eqn:sum_rule}). Parameters defining the functional generator of Eq.~(\ref{Eqn:gener}) for $\lambda=0.5$ used throughout the work were $N_{h}=17$, $M=2$, $N_{phmax}=15$.}
\label{fig1}
\end{figure}

For $\lambda=0.5$ and electric field $F=1/7$ the system displays nearly adiabatic Bloch oscillations in the real--time current $j(t)$ as shown in Fig.~\ref{fig1}(a), which are consistent with the bipolaron ground state dispersion. 
Weak damping due to inelastic scattering on phonons is observed, that is in turn reflected in slow increase in the total system energy $\Delta h(t)$ in Fig.~\ref{fig1}(c).  For larger electric field $F=1/3$ the reminiscence of the Bloch oscillations are still seen, however the  current remains positive at all $t>0$. From the available data we estimate the threshold field $F_{th}\sim 0.2$.
Although for $\lambda=0.5$ we have not strictly reached the quasistationary state since oscillations in current remain well pronounced, the total energy $\Delta h(t)$ shows increase in time which is approximately linear. The increase of $\Delta h(t)$ allows us to use a linear fit according to Eq.~(\ref{hQS}) and to calculate the value of quasistationary current $\bar\jmath$. The corresponding current--field characteristics and the influence of model parameters on $\bar\jmath$ will be discussed in Sec.~\ref{sec4}.

We also observe a quasi--linear increase of the total phonon quanta in the system $\langle n_{ph} \rangle(t)$, see Fig.~\ref{fig1}(b), and we find that $\Delta \dot h(t) \gtrsim \omega_{0} d \langle n_{ph}\rangle/dt$. 
While the equality would mean that the entire energy absorbed from the electric field is absorbed by the lattice, 
the  inequality signifies  that there exists  excess energy that is responsible for the  dissociation of the bipolaron. 
Further insight into this process can be obtained by calculating the average particle distance  $d(t)$ defined as
\begin{equation}
d(t)=\sqrt{\sum_{i,j}\ave{n_{i} n_{i+j}}j^2},	
\label{avedist}
\end{equation}
shown in  Fig.~\ref{fig1}(d). 
We detect an overall increase of $d(t)$ for both values of $F$, signaling the dissociation of the bipolaron. We discuss the process of dissociation in more detail in Sec.~\ref{a} and Sec.~\ref{b}.

\begin{figure}
\includegraphics[width=0.45\textwidth]{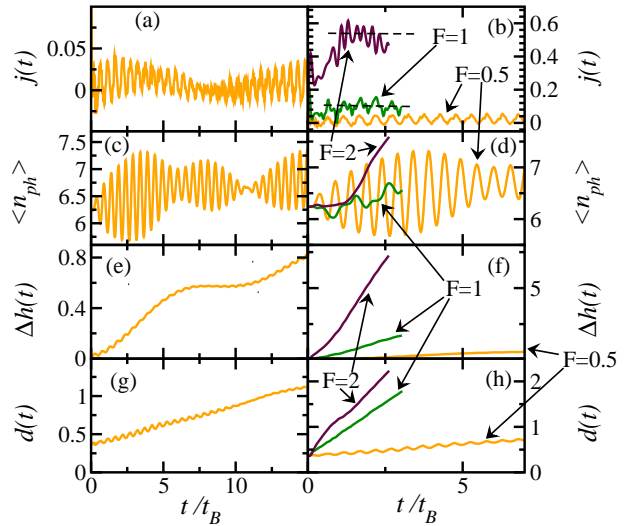}
\caption{(Color online)
$j(t)$, $\langle n_{ph} \rangle(t)$, $\Delta h(t)$ and $d(t)$ vs $t/t_{B}$ for $\lambda=0.9$.
Left column: $F=0.5$ and a maximal $t/t_B=15$.
Right column: $F=0.5$, $1$ and $2$ with a maximal $t/t_B=8$.
Thin horizontal line in (b) indicate the quasistationary current $\bar\jmath$.
Thin lines in (e) and (f) represent a linear dependence of $\Delta h(t)$, which indicates  the quasistationary state.
Parameters defining the functional generator of Eq.~(\ref{Eqn:gener}) for $\lambda=0.9$ used throughout the work were $N_{h}=13$, $M=3$ and $N_{phmax}=15$.}
\label{fig2}
\end{figure}

In the strong--coupling regime at $\lambda=0.9$ we calculate the real--time response of the system at $F=0.5 \approx F_{th}$, shown in Figs.~\ref{fig2}(a),~\ref{fig2}(c),~\ref{fig2}(e) and~\ref{fig2}(g).
Remarkably, beside the previously mentioned effects, we observe beats on a longer time--scale.
We discuss their origin in Sec.~\ref{sec5}.
If the values of electric field are increased above $F\approx F_{th}$, the beats become less pronounced.
For $F=1$ and $F=2$ we obtain, after initial oscillations, a well defined quasistationary current $j(t)$ and a nearly linear time--dependence of $\langle n_{ph} \rangle(t)$ and $\Delta h(t)$ characteristic for the dissipative regime, see Figs.~\ref{fig2}(b),~\ref{fig2}(d) and~\ref{fig2}(f).

\subsection{Bipolaron dissociation} \label{a}

The central goal of our work is to understand whether in the long--time limit the bipolaron remains bound (in some parameters regimes) as it travels under the influence of the constant electric field, or it dissociates into two separate propagating polarons.
The most interesting is the regime of strong EP interaction where the bipolaron binding energy $\delta$ is larger than the phonon energy $\omega_0$.
Such case is shown in Figs.~\ref{fig2}(g) and~\ref{fig2}(h) for $\lambda=0.9$.
The  average distance $d(t)$ is steadily increasing even in the near-adiabatic regime,
and the slope of the linear increase of $d(t)$ is enhanced in the dissipative regime, as seen for $F=1$ and $2$ in  Fig.~\ref{fig2}(h).

We should stress that the overall increase of $d(t=t_{\mathrm {max}})$ after the maximal propagation time relative to the initial ground state distance $d(t=0)$  is more than 6--fold in both cases, {\it i.e.}, at $F=1$ and $F=2$.  Even though  the maximal propagation time is in our time-evolution limited with the maximal allowed size of the Hilbert space, our results  strongly suggest  that in the limit when  $t\rightarrow\infty$ the bipolaron dissociates  into two separate polarons. 

Additional information about the bipolaron dissociation can be obtained if $d(t)$ is compared to the average distance traveled by the center of mass of two particles $\Delta x(t)$, where

\begin{equation}
 \Delta x(t)=\frac{\Delta h(t)}{2F}.
\end{equation}

Using $\Delta x(t)$ it is convenient to define a ratio

 \begin{equation} \label{eta}
\eta(t)= \frac{ \Delta x(t)}{\Delta d(t)},
 \end{equation}
which represents the ratio between the average distance $\Delta x(t)$ traveled by two particles along the field direction, and the relative increase of the distance $d(t)$ between the particles defined as $\Delta d(t) = d(t) - d(t=0)$.

\begin{figure}
   \centering
    \begin{tabular}{c}
    \includegraphics[width=1.0\linewidth]{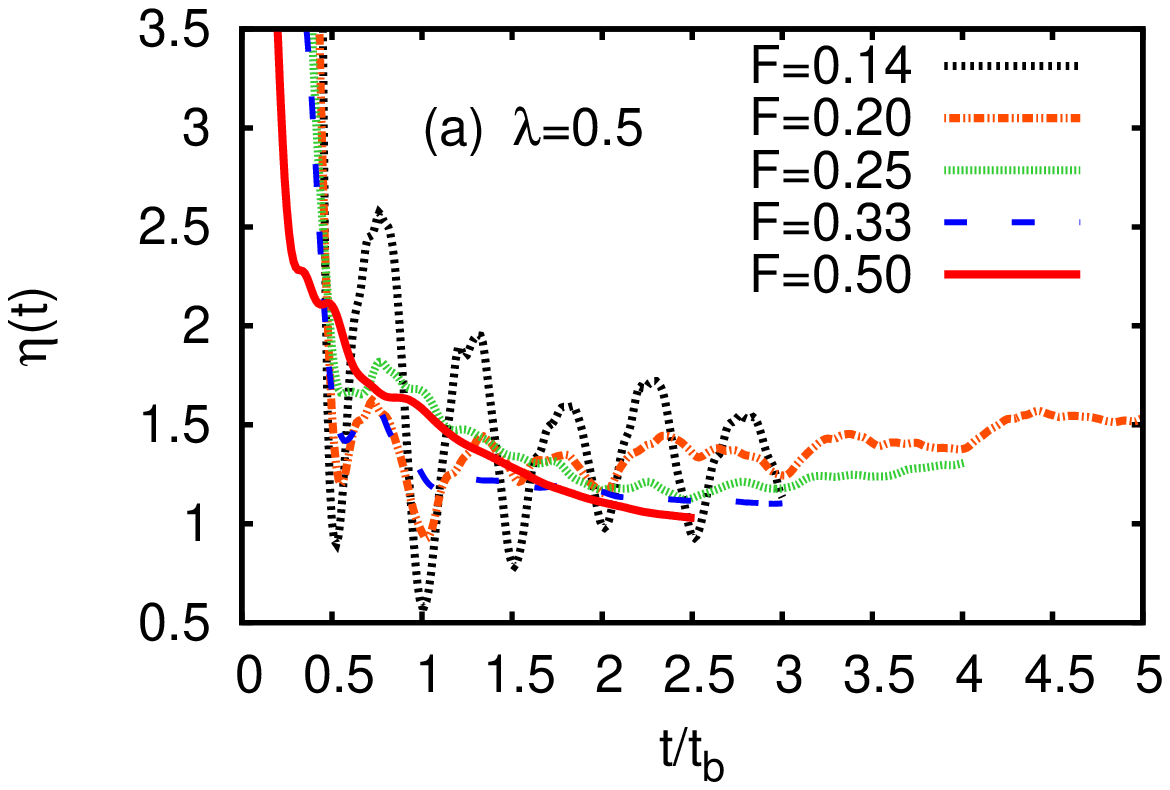} \\
    \includegraphics[width=1.0\linewidth]{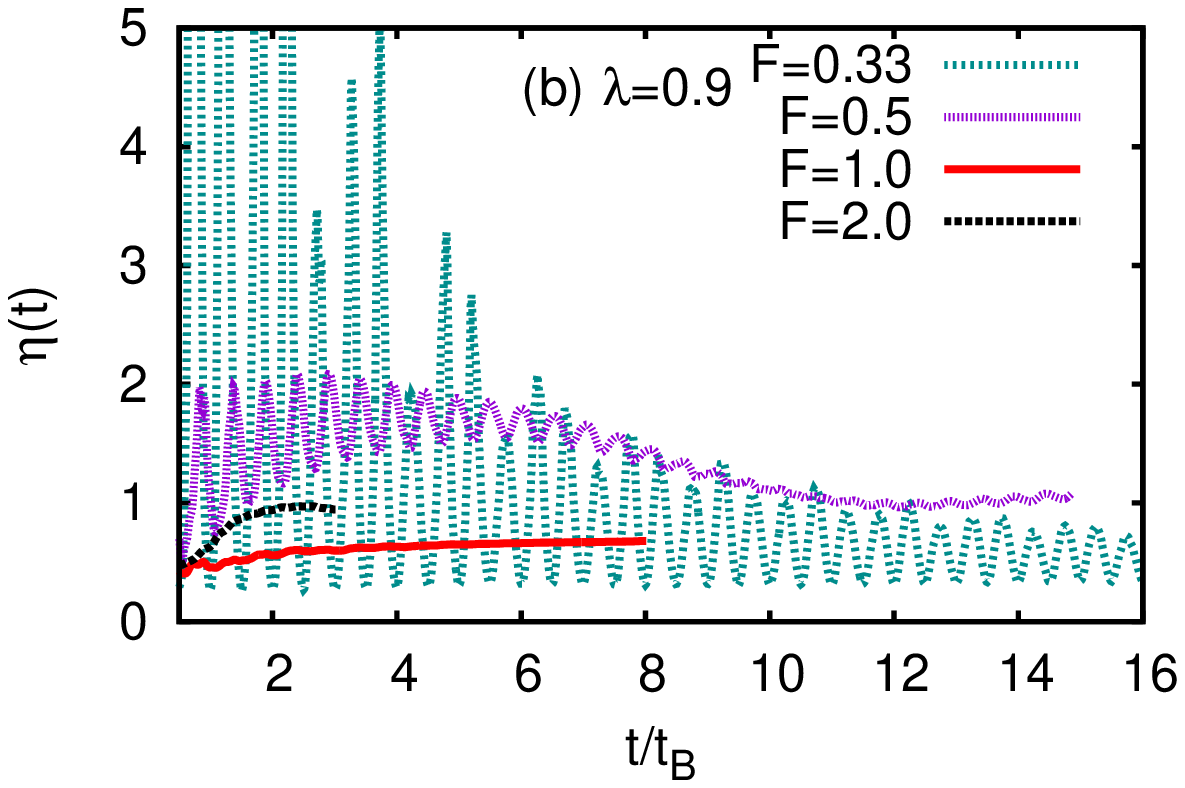} 
    \end{tabular}
    \caption{The ratio $\eta (t)$ as defined in Eq.~(\ref{eta}), vs $t/t_B$. Upper panel and lower panel correspond to $\lambda=0.5$ and $0.9$, respectively.}
    \label{fig:quotient}
\end{figure}

If the bipolaron propagated without dissociating into separate polarons, we would expect a linear increase of $\eta(t)$ in the quasistationary state.   
Instead, results for both  $\lambda=0.5$ and $0.9$ displayed in Fig.~\ref{fig:quotient} show a clear tendency toward a constant value of the order of one, independently on the strength of $F$. For small $F \sim F_{th}$,  $\eta(t)$ shows damped Bloch oscillations  while for larger $F$ the approach toward a constant is rather monotonous.
This results suggest that the dissociation of  the bipolaron   and the appearance of the quasistationary current  emerge  simultaneously as the system evolves from a transient regime to the quasistationary state.

\subsection{Phonon correlation function} \label{b}

\begin{figure}
\includegraphics[width=0.47\textwidth]{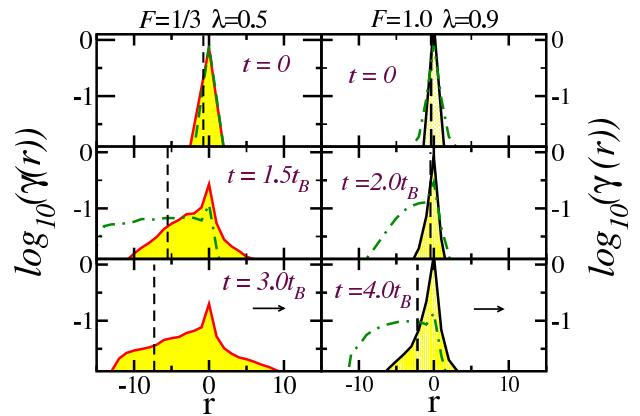}
\caption{(Color online)
$log_{10}(\gamma(r))$ for $\lambda=0.5$ $(0.9)$ and $F=1/3$ $(1)$ computed at different times.
Note that the logarithm of the correlation functions is presented.
The vertical dashed lines represent the average distance between the electrons $d$ at a certain time.
The arrow represents the direction of the moving bipolaron.
Green dashed--dotted lines represent the corresponding $log_{10}(\gamma(r))$ of the Holstein polaron model.~\cite{holE}
We show $log_{10}(\gamma(r))$ for $|r|\ll N_h$ where results for different system sizes show good convergency.
}
\label{fig3}
\end{figure}

We also compute the average number of phonon quanta located at a given distance $r$ from (both) electrons. To study this effect we define
\begin{equation}
  \gamma(r)=\frac{1}{2\ave{n_{ph}}} \langle \sum_{i,\sigma} n_{i,\sigma}a^{\dagger}_{i+r}a_{i+r}\rangle,
\label{Eqn:gamma}
\end{equation}
fulfilling the sum rule $\sum_{r} \gamma (r)=1.$
Correlations function $\gamma (r)$ for $\lambda=0.5$ $(0.9)$ and $F=1/3$ $(1)$ are displayed  in Fig.~\ref{fig3}, 
giving additional  insight into the dynamics of the moving bipolaron. The most characteristic feature is  a pronounced asymmetry of $\gamma(r)$ with respect to the electron positions at $r=0$ that grows with time.  The asymmetry  emerges due to phonon excitations  extending behind the moving particles, 
which contain the excess energy absorbed from the external electric field. Here we note that the electron is moving in the direction of $r>0$.  This effect is similar to  the polaron case.~\cite{holE} 
The second feature is the  increased amount of  phonon excitations  in the forward direction, which arises due to two distinct contributions. 
While the first contribution is due to damped Bloch oscillations analogous to the polaron propagation,~\cite{holE} the second contribution is entirely due to a two--particle effect, in which case there is always one particle traveling ahead in the electric field.  It is the second particle which follows the first one   that in turn  detects phonon excitations left by the first particle, thus generating extra weight of $\gamma(r)$ in the region  $r>0$.  
Therefore, to obtain the complete explanation of the increasing $\gamma(r)$ in the forward direction one has to take also into account the increasing  average distance between  two particles. 

For comparison we added results for the polaron case, presented by  dashed--dotted lines in Fig.~\ref{fig3}, which show  a farther extend of $\gamma(r)$ behind the moving polaron  and a smaller reach of $\gamma(r)$ in the forward direction, the latter being consistent with the argument given above.
A shorter phonon disturbance behind the traveling bipolaron  indicates  a smaller center of mass velocity  in comparison to the  polaron. 
We further elaborate on this issue in the following section. 

\section{Quasistationary current} \label{sec4}

\begin{figure}
   \centering
   \begin{tabular}{cc}
     \includegraphics[width=0.5\linewidth]{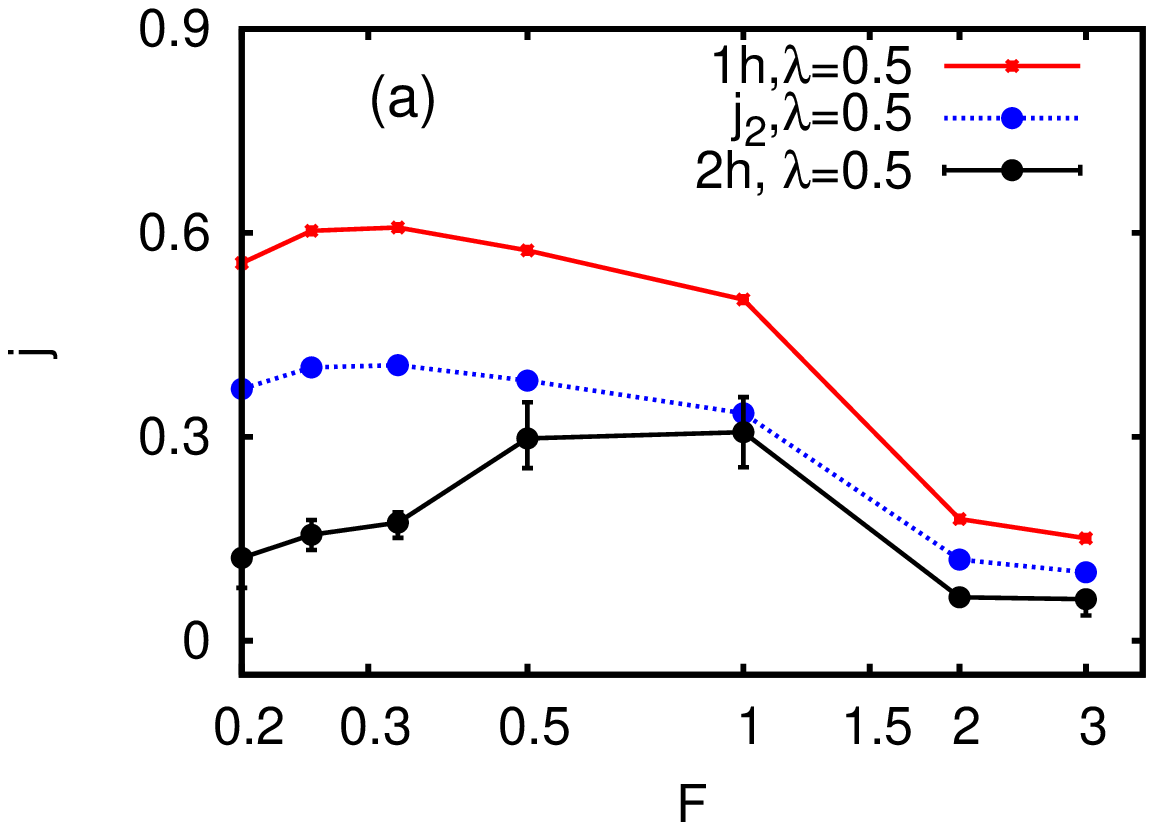}
     \includegraphics[width=0.5\linewidth]{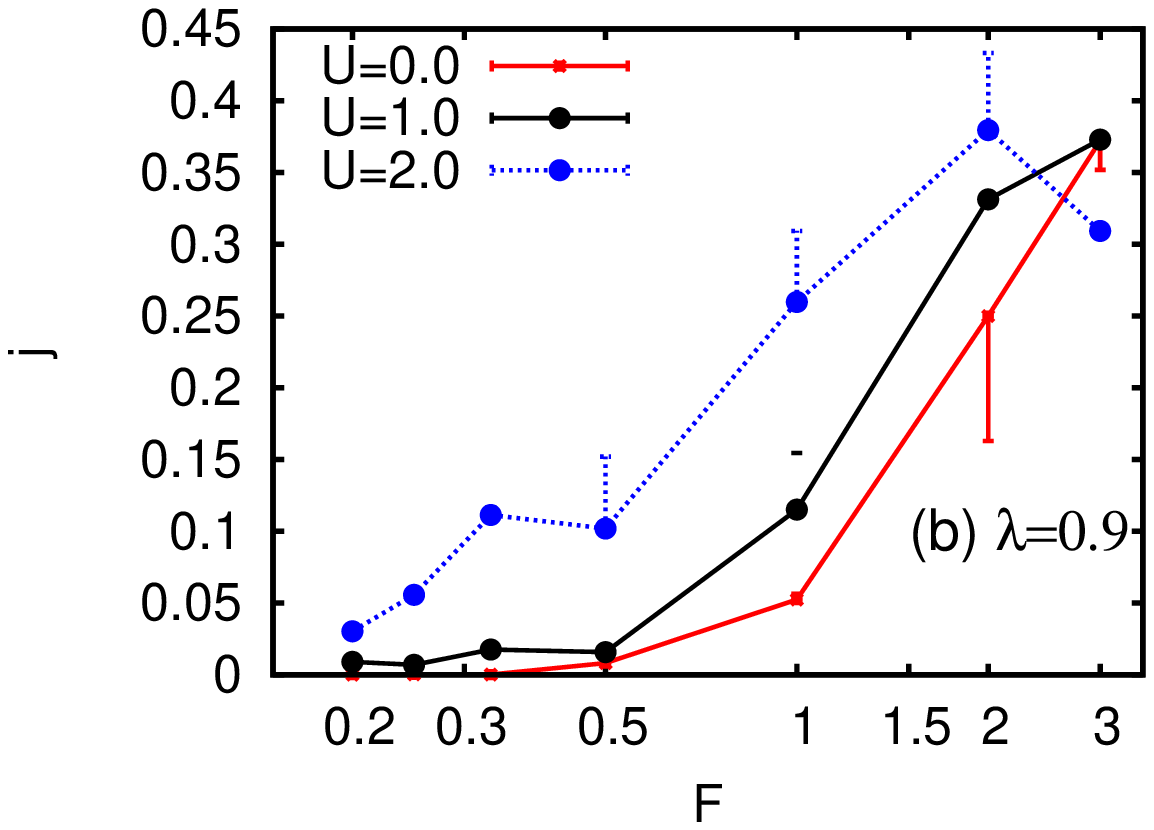} \\
     \includegraphics[width=0.6\linewidth]{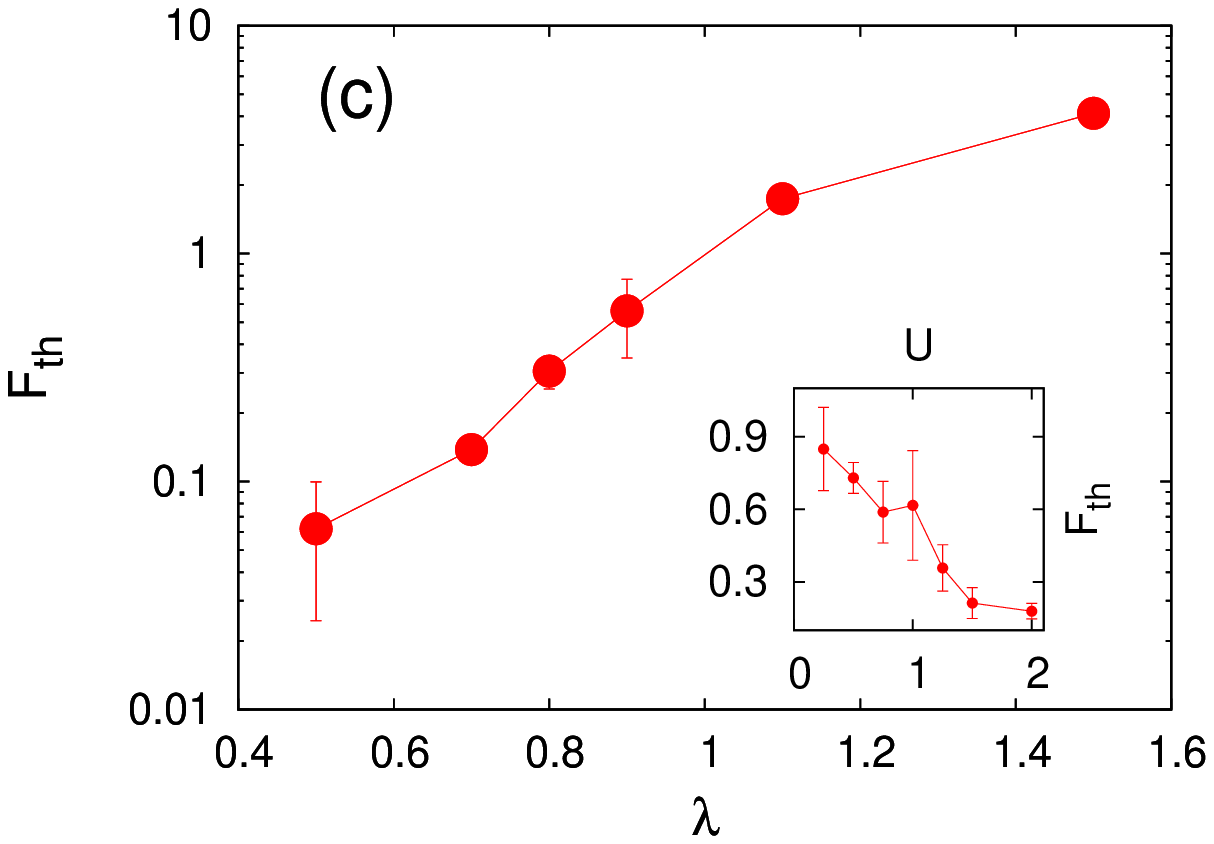}
   \end{tabular}
\caption{
 (Color online) 
 (a) Quasistationary current $\bar\jmath$ vs electric field $F$ at $\lambda=0.5$ for a polaron (1h) and a bipolaron (2h).
 The bipolaron quasistationary current estimated from the polaron quasistationary current is denoted by $\bar\jmath_2$ (see text for details).
 The uncertainty of $\bar\jmath$ is represented with error bars and is a consequence of transient region; see discussion in text. (b) Quasistationary current $ \bar\jmath$ vs electric field $F$ for $\lambda=0.9$ and different values of the Hubbard repulsion $U=0$, $U=1$, and $U=2$. The uncertainty for $\bar\jmath$  is represented with error bars and is again a consequence of transient region.
 (c) The threshold electric field $F_{th}$ as a function of the electron-phonon coupling $\lambda$ for $U=0$. The inset represents dependence of the  threshold electric field $F_{th}$ as a function of electron-electron repulsion $U$ at fixed $\lambda=0.9$.
 }
 \label{fig:eq_curr}
 \end{figure}

We next present results for the quasistationary current $\bar\jmath$, its dependence on $F$ and $U$, and comparison with the polaron case. In Fig.~\ref{fig:eq_curr}(a) we display  the current--field characteristics for $\lambda=0.5$. We have limited our calculations to commensurate values of $F=\omega_{0}/n$  for $F < \omega_{0}$ and $ F=n \omega_{0}$ for $F>\omega_{0}$ and  $n$ is integer.  
The current in the bipolaron case is in this case smaller than the current in the polaron case. This is a consequence of the fact that in the former case  one of the two electrons is traveling in the wake of the other electron where the wake  consists of the finite number of phonon excitations.  The intuitive explanation is that in this region  the effective temperature is elevated  that in turn leads to the lowering of the current. To check this assumption we performed a numerical test, where we have propagated the  polaron from the ground state until it has reached the quasistationary state. Then we changed the direction of the electric field, which prompted   the polaron to propagate backwards into the phonon rich area. After initial oscillations the polaron has  reached the quasistationary state with a lower current, {\it i.e.}, for $\lambda=0.5$ the current in the phonon rich area was roughly half of the value, obtained  when it traveled into the undisturbed  region. 

Assuming that $\bar\jmath$ is reduced in a two--particle case due to the arguments given above, and setting $\eta (t) \approx 1$ in the quasistationary state for $\lambda = 0.5$ (see Fig.~\ref{fig:quotient}(a)), it is possible to estimate the $\bar\jmath$--$F$ characteristics of a driven bipolaron from a polaron $\bar\jmath$--$F$ characteristics.
If $\Delta \dot x(t) = v_1 (1+a)/2$, where $v_1$ is the polaron velocity in the quasistationary state and $a<1$ is the renormalization of the second particle's velocity, then $\Delta \dot d(t) = v_1 (1-a)$.  Using a rough estimate  $\eta\approx 1$ yields the renormalization factor $a\approx 1/3$.
This enables us to estimate the two--particle current as $\bar\jmath_2 (F) \approx \frac{2}{3} \bar\jmath_1 (F)$.
Indeed, these results shown in Fig.~\ref{fig:eq_curr}(a) are in reasonable agreement for $F\gtrsim 0.5$ with the quasistationary current of a driven bipolaron calculated from the real--time dynamics. Using these considerations we can understand almost linear dependence of $d(t)$ in Fig.~\ref{fig1}(d) and \ref{fig2}(h), since the  particle the follows the first one moves with a renormalized but a constant velocity.

At larger lambda $\lambda=0.9$ as shown in Fig.~\ref{fig:eq_curr}(b), the bipolaron remains in the near adiabatic regime up to  the  threshold field  $F_{th}\approx 0.5$ that is much larger than $F_{th}\sim 0.1$ for the polaron case.  
The dependence of $F_{th}$ on model parameters  is better understood when loosely applying a simple Landau-Zener (LZ) formalism~\cite{landau,zener} to a multi-band energy spectrum of the bipolaron, which gives~\cite{oka03,holE}
\begin{equation} \label{lz}
F_{th}^{LZ} = \frac{(\Delta /2)^2}{v},
\end{equation} 
where $v$ is the group velocity of the quasiparticle.
Much larger $F_{th}$ in the case of bipolaron is attributed to the much narrower bipolaron bandwidth  $W_{bi}$ in comparison to the polaron $W_{pol}$, which leads to a larger gap $\Delta$ and a lower group velocity $v$ of the quasiparticle. We can estimate the scaling of $F_{th}^{LZ}$ for bipolaron in the strong coupling limit. The gap  is $\Delta \approx \omega_{0}(1-4t_0 \exp(-8 \lambda t_0 / \omega_{0}))$ and $v$ is approximated as maximal value of $d E_{bi}/d k$, which in the asymptotic expansion scales as $v \approx 2t_0\exp(-8 \lambda t_0/\omega_{0})$. Therefore to the leading order in $\lambda$, $F_{th}$ scales as $F_{th}\approx (\omega_0^2/t_0)\exp(8 \lambda t_0 / \omega_{0})$.
To extract $F_{th}$ from $\bar\jmath$ we used 
\begin{equation}
\bar \jmath(F)=\sigma_{0} F e^{- \pi F_{th}/F}, 
\end{equation}
where $\sigma_{0}$ and $F_{th}$ are fitting parameters.

Nearly linear  scaling of $\log(F_{th})$ with $\lambda$ is represented in Fig.~\ref{fig:eq_curr}(c). Here we should emphasize that in the leading order of $\lambda$, $F_{th}$ is increased predominantly  due to the narrowing of the bandwidth and consequently reducing of the group velocity $v$ in Eq.~(\ref{lz}).

In Fig.~\ref{fig:eq_curr}(b) we also present the influence of the Hubbard repulsion $U$ on the quasistationary current.
The most important effect is the increase of $\bar\jmath$ with increasing $U$ for $F\lesssim 2$.
In several recent studies of interacting systems driven out of equilibrium, an increase of a quasistationary current due to stronger correlations has been observed.~\cite{freericks2008,hm10,tjE}
In the case of the  Hubbard--Holstein bipolaron the effect of $U$ is to reduce the double occupancy and consequently the binding energy $\delta$. Moreover, in the framework of the strong EP coupling limit, $U$ reduces the effective mass and thus  increases the bandwidth,\cite{bonca00} which in turn leads to a decrease of the gap in the bipolaron excitation spectrum.
These combined effects lead to a predominantly linear decrease of $F_{th}$ with increasing $U$ in the regime $U\lesssim 1$, as represented in the inset of Fig.~\ref{fig:eq_curr}(c).  Nevertheless, $F_{th}$ tends to a constant value $F_{th}\sim 0.2$ already around $U\sim 1.5$ while the bipolaron in the ground state remains bound up to $U_c\sim 3$, see Ref.~\cite{bonca00}.  We also note that a similar threshold value $F_{th}\sim 0.1-0.2$ is found in the case of a single polaron.~\cite{holE} The bipolaron thus reaches threshold field values of a single polaron even in the regime when $U<U_c$.

 Here we should note that in case of $\lambda=0.9$ we were unable to reach (for all electric fields) a fully dissociated state of two separated polarons due to a long dissociation time and a limited variational Hilbert space. Even in this transient regime system shows clear characteristic features of quasistationary dependence, {\it i.e.}, linear dependence of energy with time, see Fig.~\ref{fig2}(f).  $\bar\jmath$--$F$ dependence for $\lambda=0.9$ represented in Fig.~\ref{fig:eq_curr}(b) is calculated in this regime, where the effects of the correlations between particles are still strong and we would expect that after the transient regime the system would enter a new regime, when bipolaron would be well dissociated and dynamics would be more similar to the $\lambda=0.5$ case.

\section{Fourier analysis} \label{sec5}
\begin{figure}
\includegraphics[width=0.47\textwidth]{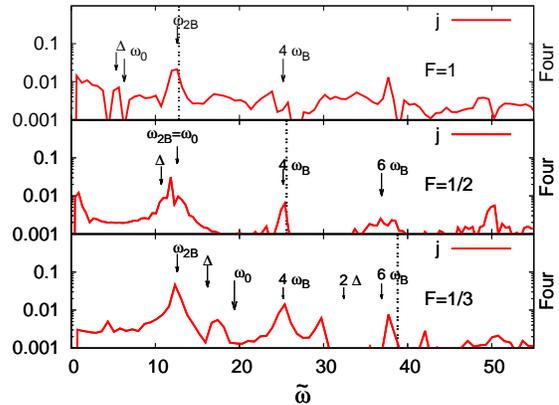}
\caption{(Color online) Normalized Fourier transform of the real--time current $j(t)$, where linear and quadratic terms have been extracted. We set electron-phonon coupling $\lambda=0.9$ and electric fields $F=1/3,1/2,1$. Arrows mark the angular frequencies corresponding to the distinct energy scales of the system. The horizontal dashed line represent the frequency corresponding  to the bipolaron binding energy. 
}
\label{fig:fourier}
\end{figure}

Following the idea of Refs.~\cite{claro2003a,khomeriki2010}  we perform the Fourier analysis of $j(t)$ in order to extract additional information about  the dynamics of the driven system. In doing so we first subtract the linear and quadratic terms from $j(t)$ data in all cases where applicable. 
In Fig.~\ref{fig:fourier} we represent the normalized Fourier transform (the integral over whole spectrum is 1) of the real--time current for $\lambda=0.9$ and different electric fields. We are measuring time in units of Bloch time  $2 \pi / F$ and the angular frequency is accordingly given by $\tilde \omega = \omega*t_B$. 

In the (nearly) adiabatic regime, for  $F=1/3$, the bipolaron exhibits Bloch oscillations, with the doubled Bloch frequency $\tilde \omega_{2B}=2\tilde \omega_B=4\pi$. This is reflected in a well pronounced  peak in the Fourier spectrum of $j(t)$ at $\tilde \omega_{2B} $. We as well observe well separated peaks at higher harmonics of $\tilde \omega_{2B}$, however,  with decreasing  amplitude. 

In the strong--coupling picture processes responsible for such peaks are diagonal transitions,\cite{mahan} {\it i.e.}, processes where bipolaron exhibits coherent propagation through the lattice. For example,  if the bipolaron jumps as a bound pair for one site along the field direction, the  energy difference between the initial and the final state is $2 F$. The related frequency for such a process is $\omega=2 F$ or $\tilde \omega= 4 \pi$. Jumps for more sites represent  higher order processes, which are responsible for generation of higher harmonics. Other peaks are related to different energy scales in the system, such as  $\omega_{0}$ and $\Delta$ as indicated  in Fig.~\ref{fig:fourier}.

For $F=1/2$ we noticed pronounced  beats in different observables, see Figs.~\ref{fig2}(a), (c), (e) and (g).
In Fig.~\ref{fig:fourier} at $F=1/2$ the most pronounced peak corresponds again to $\tilde \omega= 4 \pi$, however, it this case the  bipolaron Bloch frequency $\tilde \omega_{2B}$ matches the phonon frequency $\omega_0$.
This is in contrast to the case at $F=1/3$ where we notice no such accidental degeneracies, and no beats appear in $j(t)$.  Another condition for the appearance of the beats is that the electric field is in the vicinity of the threshold field, {\it i.e.}, $F\sim F_{th}$, since at larger $F$ increased energy dissipation to phonons overdamps  Bloch oscillations. Pronounced beats in $j(t)$ were observed as well by other authors exploring different  driven models.~\cite{claro2003a,freericks2008}

For $F=1$ the system reaches the quasistationary state in a time $t\lesssim t_B$ with a nearly linear total energy increase  $\Delta h(t)$, see Fig.~\ref{fig2}(f), and a rapid increase of $d(t)$, see Fig.~\ref{fig2}(h).
For this reason the Fourier spectrum in  Fig.~\ref{fig:fourier} of $j(t)$ shows less pronounced, broader  peaks.  We note that the frequency corresponding to the bipolaron  binding energy $\delta$ as marked with the dashed line  is comparable to $\tilde  \omega_{2B}$.  

\begin{figure}
\includegraphics[width=0.47\textwidth]{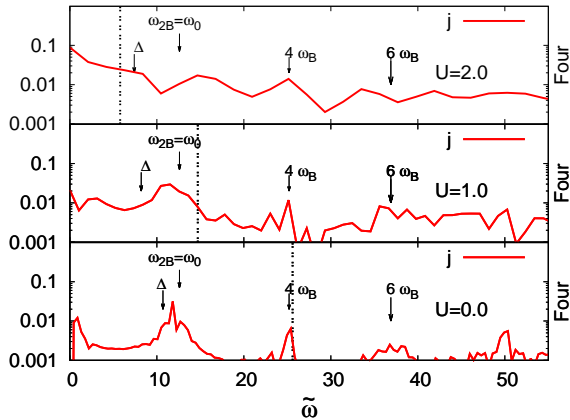}
\caption{(Color online) 
Normalized Fourier transform of the real--time current $j(t)$, where linear and quadratic terms have been extracted, for different values of Hubbard repulsion $U$. We set $\lambda=0.9$ and $F=0.5$. Arrows mark the angular frequencies corresponding to the distinct energy scales of the system. The horizontal dashed line represent the frequency corresponding  to the bipolaron binding energy. 
%
}
\label{fig:fourier2}
\end{figure}

The influence of the Hubbard repulsion on the dynamics of $j(t)$ is presented via the Fourier transform in  Fig.~\ref{fig:fourier2} for the  case of $\lambda=0.9$ and  $F=1/2$. The spectrum   at finite  $U=1.0$ shows broader peaks near the  characteristic frequencies in comparison to  $U=0$ results. The effect of increasing $U$ is in agreement with the considerations given is Sec.~\ref{sec4}.
We again note that the spectrum is broadened when the frequency corresponding to $\delta$ becomes comparable to $\omega_{2 B}$.
Despite much broader spectrum at $U=1$, $\bar\jmath$ remains small, comparable to the $U=0$ case. At larger $U=2$, the peaks in the spectrum become indistinguishable, while simultaneously a significant quasistationary current appears, as seen in  Fig.~\ref{fig:eq_curr}(b). 

\section{Conclusions} \label{sec6}

We have studied  the time-evolution  of the Holstein-Hubbard bipolaron from its ground state at zero temperature  after a  constant electric field has been  switched on at $t=0$.  Using an efficient variational method, defined on an infinite one-dimensional chain, we time-evolved the many--body Schr\"odinger equation while preserving the full quantum nature of the problem, until the system has reached a quasistationary state. 
In the limit of small electric field the bipolaron evolves along the  quasiparticle band while the current shows characteristic  Bloch oscillations. In this regime the propagation is adiabatic, there is no increase of the total energy, the net current remains zero  and the bipolaron remains bound. 
 At larger electric fields we detect an overall  increase of the total energy, a finite net current appears and the system enters the dissipative regime. Due to dissipation of the gained potential energy into lattice vibrations, Bloch oscillations in this regime become damped. After a transient time the system enters a quasistationary state with a constant current. We note that there is no clear distinction between the adiabatic and dissipative regime. While  a true adiabatic regime exists only in the limit when $F\to 0$, a sizable net current signaling the dissipative regime appears for $F\gtrsim F_{th}$. 

The focal result of our work is that in the dissipative regime  the bipolaron dissociates into two separate polarons. By examining different parameter regimes we realized that the appearance of a finite quasistationary current is closely connected with the dissociation  of the bipolaron. Even though our calculations were  limited to large $F$ and short propagation times, our results strongly suggest that bipolaron has a finite lifetime as long as the system displays a non-zero $\bar\jmath$. This hypothesis is supported by the calculation of $\eta(t)$ that in the large-time limit tends to a constant, largely independent of $F$. 
This result demonstrates   that the dissociation   and the appearance of the quasistationary current emerge  simultaneously as the system evolves from a transient regime to the quasistationary state. 

Here we should comment that due to dispersionless phonons the bipolaron can gain energy when two particles occupy the same site, but to achieve a non-zero quasistationary current charged particles have to start hopping along the direction of electric field, and in doing so  breaking the bound pair. In Ref. \cite{kuroda1985} the Green function study at finite temperature (without electric field) calculated the bipolaron dissociation time, which goes to zero for dispersionless phonons, while this is not the case for phonons with dispersion. For the problem studied here this may indicate that phonons with dispersion would not always lead to dissociation of bipolaron as soon as a finite quasistationary current is obtained. Further research in this field is necessary to resolve the issue. 

A linear time--dependence of a real--space expansion of particle densities or related quantities has been observed in several recent nonequilibrium problems.~\cite{langer11a,langer11b}
Even though we study a two--particle problem on an infinite lattice and we do not deal with a well--defined thermalization, we can consider a cloud of emitted phonon excitations as a subsystem with an elevated effective temperature. The quantum Monte Carlo study~\cite{hohenadler05b} of the thermal dissociation of the  Hubbard--Holstein bipolaron provides evidence that a bipolaron dissociates as  the temperature becomes comparable to the binding energy $\delta$. In comparison, our results indicate that the bipolaron dissociates as soon as a finite net current appears around $F \sim F_{th}$. This occurs even in the case of $\delta > \omega_0$, which is realized for $\lambda=0.9$ in our calculations, see Fig.~\ref{fig2}(h).
We have also  checked the time dependence of systems at even larger electron-phonon coupling,  up to $\lambda\le1.5$.  With increasing $\lambda$ the  qualitative behavior remains the same, the bipolaron as well dissociates with the dissociaton time that is increasing with $\lambda$ and   the  threshold field is also roughly exponentially increased as shown in Fig.~\ref{fig:eq_curr}(c).

Yet another noteworthy finding reveals that the  quasistationary current per particle is in case of the dissociated bipolaron smaller than in the single polaron case since the velocity of one particle is diminished due to the scattering on phonons generated by the motion of the other particle.  Comparison of the bipolaron $\bar\jmath$--$F$ characteristics at $\lambda=0.9$ with the polaron case (not presented) shows that a quasistationary current of  bipolaron is not always lower than that of the polaron, which is possibly a consequence of remnant correlations between particles due to the incomplete dissociation of the bipolaron in the transient regime.

To summarize, we have demonstrated the dissociation of the bipolaron under the influence of the electric field. This result predicts that as long as a given dilute  system of bipolarons can be described by the short range Hubbard model in 1D and the on-site coupling term between the charge density and  the dispersionless phonons, bipolarons do not carry electric current, instead they  dissociate into separate polarons as soon as the electric field is strong enough to yield a finite current.


\acknowledgements
L.V., J.B. and D.G. acknowledge stimulating discussions with M. Mierzejewski, V. Zlati\v c and T. Tohyama.  We thank also P. Prelov\v sek for his inspiring remark. This work has been support by the Program P1-0044 of the Slovenian Research Agency (ARRS),  REIMEI project, JAEA, Japan, and CINT user program, Los Alamos National Laboratory, NM USA.

\end{document}